\documentclass[12pt]{article}
\usepackage{amssymb}
\usepackage{graphicx}
\def\be{\begin{equation}}
\def\ee{\end{equation}}

\begin{document}
\begin{center}
\hfill \vbox{
\hbox{DFCAL-TH 03/1}
\hbox{February 2003}}
\vskip 0.8cm
{\Large\bf The spectrum of screening masses near $T_c$: predictions from   
universality.}\\ 
\end{center}
\vskip 0.3cm
\centerline{R. Fiore$^{a~\dagger}$,  
A. Papa$^{a~\dagger}$ and P. Provero$^{b~\ddagger}$}
\vskip 0.3cm
\centerline{\sl $^a$ Dipartimento di Fisica, Universit\`a della Calabria}
\centerline{\sl Istituto Nazionale di Fisica Nucleare, Gruppo collegato
di Cosenza}
\centerline{\sl I--87036 Arcavacata di Rende, Cosenza, Italy}
\vskip0.2cm
\centerline{\sl $^c$ Dipartimento di Fisica
Teorica dell'Universit\`a di Torino}
\centerline{\sl Istituto Nazionale di Fisica Nucleare, Sezione di Torino}
\centerline{\sl via P. Giuria 1, I--10125 Torino, Italy}
\vskip 0.6cm
\begin{abstract}
We discuss the spectrum of screening masses in a pure gauge theory near the 
deconfinement temperature from the point of view of the dimensionally reduced 
model describing the spontaneous breaking of the center symmetry. 
Universality arguments can be used to predict the values of the mass ratios in 
the scaling region of the deconfined phase when the transition is of
second order. One such prediction   
is that the scalar sector of the screening spectrum in $SU(2)$ pure gauge 
theory contains a bound state of the fundamental excitation, corresponding 
through universality to the bound state found in the 3D Ising 
model and $\phi^4$ theory in the broken symmetry phase. A Monte Carlo 
evaluation of the screening masses in the gauge theory confirms the
validity of  
the prediction. We briefly discuss the possibility of using similar arguments 
for first order deconfinement transitions, and in particular for the
physically  
relevant case of $SU(3)$.
\end{abstract}

\vfill
\hrule
\vspace{0.3cm}
\noindent$^\dagger$ {\it e-mail address}: fiore,papa@cs.infn.it \newline
$^\ddagger$ {\it e-mail address}: provero@to.infn.it

\newpage

\section{Universality prediction for the screening \- masses}

QCD predicts the existence of a deconfinement temperature above which the 
fundamental degrees of freedom cease to be confined in color singlets and 
can propagate in isolation. This prediction is being experimentally tested 
in experiments where heavy nuclei are collided at high energies to produce 
such deconfined state of matter, the quark-gluon plasma. It is therefore   
imperative from the theoretical point of view to develop a better
understanding of the behavior of gauge theories at temperatures above the 
deconfinement transition (for a recent review see {\em e.g.} 
Ref.~\cite{Kogut:2002kj}).
\par
Since a non-zero temperature for a quantum field theory is equivalent to a 
finite size and periodic boundary conditions in Euclidean time, in many 
important respects gauge theories at high temperature can be treated as 
effectively living in one dimension less, a phenomenon generally referred to 
as dimensional reduction. More precisely, dimensional reduction can be defined 
through a specific pattern of pairwise degeneracies in the spectrum, characteristic 
of theories living in one less dimension than the original one, so that the 
occurrence of dimensional reduction can be quantitatively verified {\it e.g.} 
by Monte Carlo simulations. For $SU(2)$ and $SU(3)$ gauge theories in $3+1$ 
dimensions, this comparison was successfully performed in~\cite{Datta:1998eb}, 
where convincing evidence was provided that dimensional reduction occurs for 
temperatures all the way down to the deconfinement one.
\par
Much less is known about the actual form of the dimensionally reduced 
theory, especially for temperatures not too far from $T_c$, where perturbative 
methods are not reliable. The simplest hypothesis, that the effective
theory is the pure gauge theory with the same gauge group in one less dimension
and in the confined phase, is ruled out by numerical data on the spectrum of 
screening masses for $SU(2)$~\cite{Datta:1999yu}, while for $SU(3)$ the
agreement is better. A more sophisticated proposal was 
made in~\cite{Kajantie:1997tt}, where the effective 3D theory is taken
to be a three-dimensional gauge theory coupled to an adjoint
scalar. This effective theory has been shown to correctly reproduce the
low-lying states of the spectrum for both $SU(2)$~\cite{Hart:1999dj}
and $SU(3)$~\cite{Hart:2000ha} for temperatures down to $\sim 2T_c$.   
\par
On the other hand, in the temperature range just above $T_c$, we have the 
possibility of making a solid conjecture about the form of the dimensionally 
reduced theory, at least for the case of $SU(2)$ or, in general, for gauge 
groups such that the transition is of second order: this is the {\it scaling 
region} above the transition, where universality arguments hold and the 
effective theory can be conjectured to belong to the universality class of 
theories whose global symmetry group coincides with the center of the gauge 
group. For $SU(2)$, the relevant universality class is the one of the 
3D Ising model. This application of universality 
arguments to the deconfinement transition was introduced by Svetitsky
and Yaffe~\cite{Svetitsky:1982gs}  and many of its predictions have been
accurately verified: the values of the critical indices (the most recent
results can be found in
Refs.~\cite{Fortunato:2000hg,Fiore:2001ci,Fiore:2001pf, 
Papa:2002gt}), universal amplitude ratios~\cite{Engels:1998nv}, and correlation
functions at criticality~\cite{Gliozzi:1997yc,Fiore:1998uk}. In this
approach, the fundamental degree of freedom of the effective theory is
the Polyakov 
loop, which is the order parameter of the deconfinement transition.
\par
It is far from obvious {\it a priori} that the higher masses in the spectrum 
should be universal, since only the lowest one contributes to the free energy. 
However, it has recently been shown that for the three dimensional Ising    
universality class, in the broken symmetry phase, there is strong evidence for 
such universality: the Ising model and lattice regularized $\phi^4$ 
theory both exhibit a rich spectrum of massive excitations, and the
mass ratios coincide in the scaling region~\cite{Caselle:1999tm}. 
Particularly interesting is the existence of an excited mass in the scalar 
sector at less than twice the mass of the fundamental excitation: being 
the interaction attractive in the broken symmetry phase, it is natural to 
interpret such state as a bound state of two fundamental particles, an 
interpretation that was confirmed by studying the Bethe-Salpeter equation for 
3D $\phi^4$ theory in the continuum~\cite{Caselle:2001im}.
\par
It is therefore natural to ask whether the same spectrum characterizes $SU(2)$ 
gauge theory in the scaling region above the deconfinement temperature. This 
paper is devoted to a numerical verification of this ansatz. In particular we 
will focus on the scalar sector: we will show that a bound state of the 
fundamental screening mass can indeed be extracted from the correlation 
functions of suitably defined operators, that are constructed from Polyakov 
loops in a manner similar to that employed in~\cite{Caselle:1999tm} to study
the spectrum of the Ising model. We also evaluated the lowest screening mass 
with angular momentum 2 in the scaling region and found a value compatible with 
the one of the Ising model. Preliminary results were reported in 
Ref.~\cite{Fiore:2002fj}.
\par
In the following section we describe in detail the method we used and the 
numerical results, while Section 3 is devoted to the discussion of the results and 
some speculation about the case of $SU(3)$. 

\section{Numerical determination of the spectrum of scalar screening masses}

As a first step we considered the connected wall-wall correlation function
of the Polyakov loop on a lattice with $N_x \times N_y \times N_z \times N_t $ 
sites and lattice spacing $a$
\[
G(|z_1-z_2|) \equiv \langle \overline P(z_1) \overline P(z_2) \rangle 
- \langle \overline P(z_1) \rangle \langle \overline P(z_2) \rangle\;,
\]
where
\[
\overline P(z)\equiv\frac{1}{N_x N_y}\sum_{n_1=1}^{N_x}\sum_{n_2=1}^{N_y}
P(n_1a,n_2a,z)
\]
is the average of the Polyakov loop $P(x,y,z)\equiv \mbox{Tr} \prod_{n_4=1}
^{N_t} U_4(x,y,z, n_4a)$ over the $xy$-plane at a given $z$. The wall average 
implies the projection at zero momentum in the $xy$-plane.
The correlation function $G(|z_1-z_2|)$ takes contribution from all the 
screening masses in the $0^+$ channel, {\em i.e.} from the lowest (fundamental)
mass $m_1$, from the first excited mass $m_2$, and so on. For a periodic   
lattice, we have
\begin{eqnarray}
G(|z_1-z_2|)&=& C_0 \: e^{-m_1 L_z} \nonumber \\
&+& C_1 \biggl[e^{-m_1 |z_1-z_2|} + e^{-m_1 (L_z-|z_1-z_2|)}\biggr] 
\label{green}\\
&+& C_2  \biggl[e^{-m_2 |z_1-z_2|} + e^{-m_2 (L_z-|z_1-z_2|)}\biggr] + \ldots \;,
\nonumber
\end{eqnarray}
with $L_z=a N_z$.
Here the dots represent the contribution from higher mass excitations and from 
multi-particle cuts. It should be noticed that, since $\langle \overline P(z) 
\rangle \neq 0$ in the high temperature phase, a $z$-independent exponential
term appears in the r.h.s. of
Eq.~(\ref{green})~\cite{Montvay:1987us}. It is easy to  
verify that such term is sub-dominant with respect to the contribution of the 
fundamental mass and of the excited masses $m_i$, $i>1$ for which $m_i<2m_1$.

We discretized the theory on the lattice with the Wilson action and 
generated Monte Carlo configurations by the overheat-bath updating 
algorithm~\cite{Petronzio:gp} with Kennedy-Pendleton improvement~\cite{Kennedy:nu}. 
Measurements were taken every 10 upgradings and error analysis was performed 
by the jackknife method applied at different blocking levels.

We made simulations at two values of $\beta$ slightly above the critical value 
$\beta_c=2.29895(10)$ determined in Ref.~\cite{Engels:1998nv}: 
at $\beta=2.33$ (corresponding to $T/T_c\simeq 1.10$) we collected 1M 
measurements on a $18^2\times36\times4$ lattice and at $\beta=2.36$ 
(corresponding to $T/T_c\simeq 1.21$) we collected 1.5M measurements on a 
$18^3\times4$ lattice. 

To make sure that our Monte Carlo ensembles were not contaminated by
configurations with mixed phase, we plotted histograms of the Polyakov loop 
averaged over the lattice configuration. In all cases, histograms showed     
two non-overlapping peaks located on opposite sides with respect to the origin, 
thus indicating that no configurations with mixed phase were present in the thermal 
equilibrium ensemble.

\begin{figure}[htb]
\includegraphics[width=0.9\textwidth]{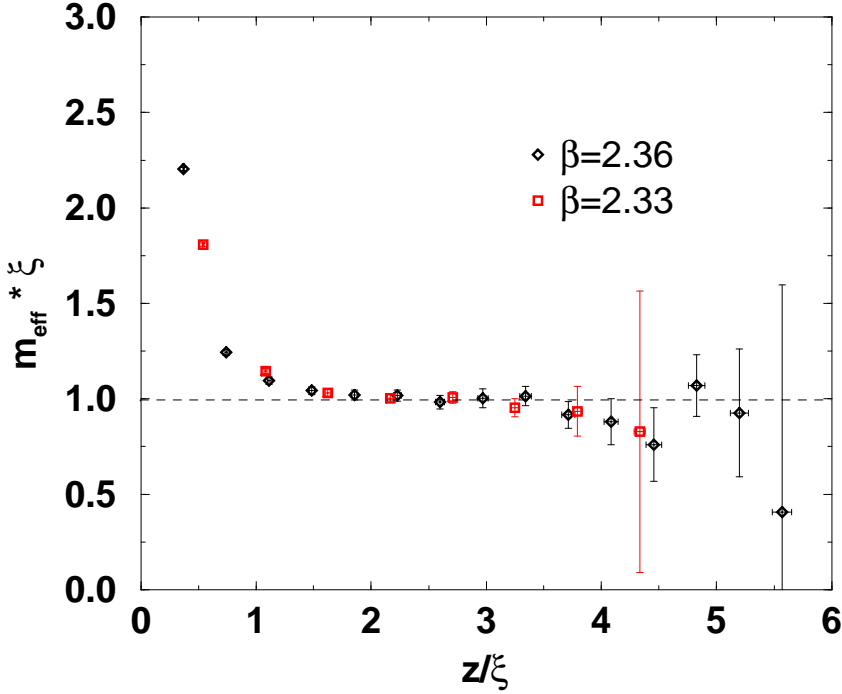}
\caption{Effective screening masses in physical units {\it vs} the 
separation between walls in physical units at $\beta=2.33$ on a 
$18^2\times36\times4$ lattice and $\beta=2.36$ on a $18^3\times4$ lattice.}
\label{scaling}
\end{figure}

We determined the effective mass through
\begin{equation}
G(z) = C_0 \: e^{-m_{\mbox{\tiny eff}} L_z}+C \biggl[e^{-m_{\mbox{\tiny eff}} z}
+ e^{-m_{\mbox{\tiny eff}} (L_z-z)}\biggr]
\label{meff}
\end{equation}
and found for both $\beta$ values that data for $m_{\mbox{\tiny eff}}$ as a 
function of $z$ reach a plateau value, corresponding to the fundamental mass 
$m_1$, as $z$ is increased towards $N_z/2$. The deviations from the plateau 
value at small $z$ can be attributed to lattice artifacts and to the possible 
effect of other physical states. 
In order to single out the latter contribution, we rescaled the values of 
$m_{\mbox{\tiny eff}}$ and $z$ by $m_1\equiv1/\xi$ and $\xi$, respectively, 
and put together on the same plot data from $\beta=2.33$
and $\beta=2.36$. As shown in Fig.~\ref{scaling}, data from the two different 
$\beta$'s fall on the same curve {\em before} reaching the plateau, 
indicating that the pre-asymptotic behavior of the effective mass is not     
a lattice artifact, but it is instead a physical effect, due to higher mass 
excitations.

Universality arguments hold in the scaling region of the deconfinement
transition: in order to check whether our choice of $\beta$ is in fact
within this region, one can compare the values of the lowest screening
mass $m_1$ at $\beta_1=2.33$ and $\beta_2=2.36$, whose ratio is
predicted by universality to scale as:
\begin{equation}
\frac{am_1(\beta_1)}{am_1(\beta_2)}\sim
\left(\frac{\beta_1-\beta_c}{\beta_2-\beta_c}\right)^\nu
\end{equation}
where $\nu\sim 0.63$ is the correlation length critical index for the
three dimensional Ising model. The values we measured\footnote{The
value quoted for $\beta_1$, which is the point where the analysis of
the whole spectrum was performed, was obtained with the variational
method described below. The value at $\beta_2$ was obtained simply
from the long-distance behavior of $m_{\mbox{\tiny eff}}$ as defined in
Eq.~(\ref{meff}). The value obtained with the latter method at
$\beta_1$ is compatible with the one quoted.} give 
\begin{equation}
\frac{am_1(\beta_1)}{am_1(\beta_2)}=\frac{0.3654(31)}{0.5431(61)}=0.673(13)
\end{equation}
to be compared to 
\begin{equation}
\left(\frac{\beta_1-\beta_c}{\beta_2-\beta_c}\right)^\nu\sim 0.65
\end{equation}
We conclude that the scaling region includes, within the accuracy of
our data, these two values of
$\beta$, and hence that such region extends at least up to temperatures
$\sim 1.2\, T_c$. On the other hand, the value of the mass at $T=2
T_c$ reported in Ref.~\cite{Datta:1999yu} is rather far from the
scaling prediction, therefore the scaling region does not extend
up to $2 T_c$.

In order to detect possible excited states, we adopted the so-called 
``variational'' method~\cite{Kronfeld:1989tb,Luscher:1990ck}. The
method consists of the following steps: 
(i) defining a basis of (wall-averaged) operators $\{O_\alpha\}$, 
(ii) computing the connected cross-correlation matrix among these operators,
\begin{equation}
C_{\alpha\beta}(z)=\langle O_\alpha(z)O_\beta(0)\rangle-
\langle O_\alpha(z)\rangle \langle O_\beta(0)\rangle\;,
\end{equation}
(iii) diagonalizing $C_{\alpha\beta}(z)$ to obtain the eigenvalues $\lambda_i(z)$
and finally (iv) extracting the masses $m_i$ through the relation 
\begin{equation}
\lambda_i(z)= c_0 +c \biggl[e^{-m_i z}+e^{-m_i (L_z-z)}\biggr]\;.
\label{lambda}
\end{equation}
The fundamental mass $m_1$ corresponds to the leading eigenvalue $\lambda_1$, 
the first excited mass $m_2$ corresponds
to the next-to-leading eigenvalue $\lambda_2$, and so on. In practical 
simulations, for each value of $z$ we have a numerical determination of      
an ``effective'' mass $m_i(z)$ through Eq.~(\ref{lambda}). As $z$ increases, 
$m_i(z)$ reaches a plateau value which we identify with the screening mass
$m_i$. In our simulations, the signal-to-noise ratio allowed to extract a 
screening mass only from the two leading eigenvalues.

The effectiveness of this method relies on the choice of a ``good'' set of 
operators; it is convenient to define operators living on different length scales
(for instance, by use of recursive definitions). Moreover, by defining 
operators with non-trivial transformation under spatial rotation, it is 
possible to look for states with non-zero angular momentum.

As a first set of operators with $0^+$ quantum numbers, we considered those
which can be built by adapting to the present case the recursive 
``smoothing'' procedure used in Ref.~\cite{Caselle:1999tm} for the 3D Ising 
model and the 3D lattice regularized $\phi^4$ theory: 
\begin{equation}
P^{(0)}(x,y,z) = P(x,y,z)\;,\;\;\;
P^{(n+1)}(x,y,z) = \mbox{sign}(u) \biggl[(1-w)|u|+w y\biggr]\;,
\label{set_I}
\end{equation}
where $u$ is the average of four nearest values of the Polyakov loop on
the $xy$-plane,
\[
u=\frac{1}{4}\biggl(P^{(n)}(x-a,y,z)+P^{(n)}(x,y-a,z) 
+P^{(n)}(x+a,y,z)+P^{(n)}(x,y+a,z)\biggr)\;,
\]
and we chose $w=0.1$ and $y=\langle P(x,y,z) \rangle$. We considered a subset 
of six such (wall-averaged) operators corresponding to $n=0$, 3, 6, 9, 12 and 15
smoothing steps. At $\beta=2.33$ ($T/T_c\simeq 1.10$) on a 
$18^2\times36\times4$ lattice, we collected $\sim$1.6M measurements.  

The relevant observable to check our ansatz is the ratio between the
first excited and the fundamental screening masses in the scalar
channel, which in the 3D Ising and $\phi^4$ theories
assumes the value 
\begin{equation}
\frac{m_2}{m_1}=1.83(3)\;.
\end{equation}
The effective values of this ratio for $SU(2)$ gauge theory are shown
in Fig. 2: while the determination of this ratio is much more
difficult than in the Ising model, the compatibility of the result
with the expectations from universality is satisfactory.
We have also observed that the excited scalar state cannot be detected
when only the first four operators of our set (0, 3, 6 and 9 smoothing levels) are
included in the analysis. This fact suggests that the first excited state in the 
$0^+$ channel couples with operators living at rather large length scales.

\begin{figure}[htb]
\includegraphics[width=0.9\textwidth]{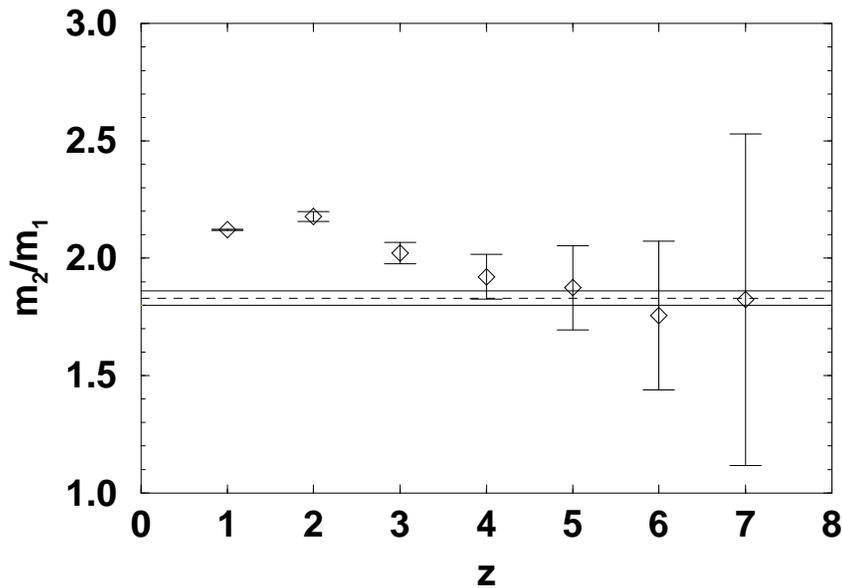}
\caption{Ratio of the first excited effective screening mass in the $0^+$ channel
to the fundamental one {\it vs} $z$ at $\beta=2.33$ on a $18^2\times36\times4$ 
lattice. The statistics of the simulation is $\sim$1.6M.}
\label{0+}
\end{figure}

\begin{figure}[htb]
\includegraphics[width=0.9\textwidth]{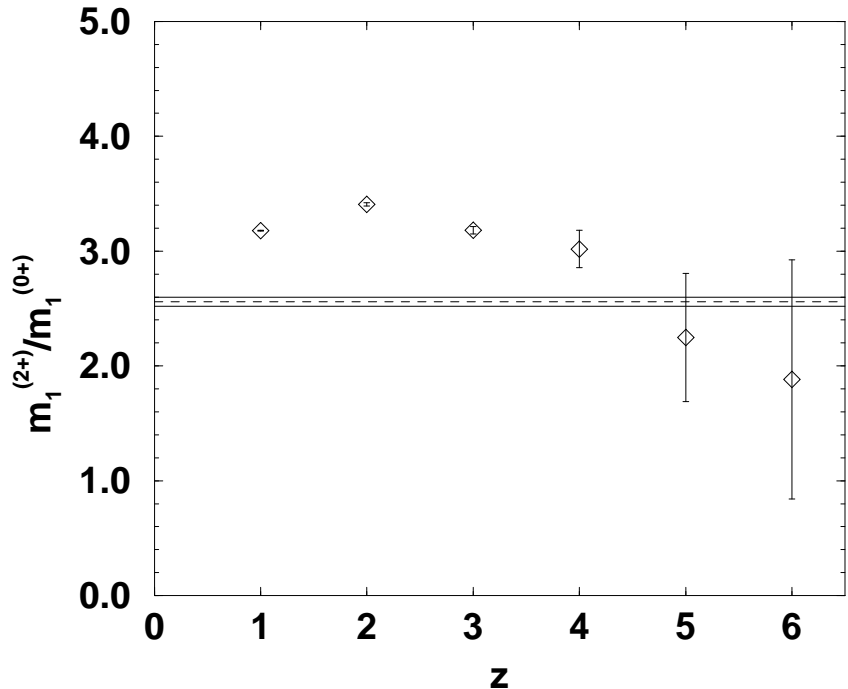}
\caption{Ratio of the fundamental effective screening mass in the $2^+$ channel
to the fundamental one in the $0^+$ channel {\it vs} $z$ at $\beta=2.33$ on 
a $18^2\times36\times4$ lattice. The statistics of the simulation is
$\sim$1.8M. The horizontal lines represent the same mass ratio for the
3D Ising model~\cite{Agostini:1996xy}, with its uncertainty.}
\label{2/0}
\end{figure}

Finally, in order to look for states with non-zero angular momentum,
we considered 
a second set of $2^+$ operators, inspired by Ref.~\cite{Agostini:1996xy}, 
\begin{eqnarray}
\overline P^{(n)}(z)&=&\frac{1}{N_x N_y}\sum_{n_1=1}^{N_x}\sum_{n_2=1}^{N_y}
P(n_1a,n_2a,z) \nonumber \\
&\times& \biggl[P(n_1a+na,n_2a,z)- P(n_1a,n_2a+na,z)\biggr]\,,
\label{set_II}
\end{eqnarray}
with $n=1,\ldots,5$. By the same procedure adopted in the case of the 
screening mass in the $0^+$ channel, we got for the fundamental
mass in the $2^+$ channel the value $m_1^{(2^+)}=0.86(21)a^{-1}$. 
Again, we studied  
the observable $m_1^{(2^+)}/m_1^{(0^+)}$, {\em i.e.} the ratio between the
fundamental mass in the $2^+$ channel, determined from this second basis of 
operators, and the fundamental mass in the $0^+$ channel, determined from the first 
basis of operators. The result of our determination, as a function of $z$, is shown 
in Fig.~(\ref{2/0}). Despite the noisiness, one can again see that the
result is compatible with the value found~\cite{Agostini:1996xy} in the 3D
Ising universality class\footnote{This mass ratio was actually
evaluated in Ref.~\cite{Agostini:1996xy} for the 3D Ising gauge model: however
duality implies~\cite{Caselle:2001im} that in the broken symmetry phase the
spectrum coincides with the one of the 3D Ising model.}
\begin{equation}
\frac{m_1^{(2^+)}}{m_1^{(0^+)}}=2.56(4)\;.
\end{equation}

\section{Discussion}
The numerical results we have obtained are in agreement with our
starting hypothesis: the
spectrum of screening masses for $SU(2)$ gauge theory near $T_c$ is
indeed the one predicted by universality arguments, {\em i.e.} it
coincides with the spectrum of the transfer matrix of the
3D Ising model in the scaling region of the
broken-symmetry phase. In particular a bound state of two elementary
screening masses can be detected. An unambiguous verification of the
coincidence of the two spectra will require both the analysis of more
states and a substantial increase in the accuracy in the masses we have
determined. 
\par
At first sight, application of these methods to the case of $SU(3)$
does not seem possible, given that the deconfinement transition is
first order. However, the transition is rather weak, both in the gauge
theory and in the 3-state Potts model that would be in the same
universality class if universality arguments applied. Therefore the
three-dimensional correlation length at criticality, 
while not diverging, assumes rather high values: hence, there exists a range 
of temperatures in which the correlation length is large in terms of lattice 
spacing, and it is not unreasonable to expect the spectra of the two theories 
to be similar in such region. We plan to investigate this issue studying
the spectrum of the three-dimensional 3-state Potts model in the broken 
symmetry phase, and comparing it to the spectrum of $SU(3)$ pure gauge theory 
near $T_c$.
\par\noindent
{\bf Acknowledgement.} We are grateful to M. Caselle for many useful
discussions.

\end{document}